\documentclass[journal,twocolumn,final]{IEEEtran}
\newcommand{\figwidth}{\columnwidth}
\newcommand{\binom}[2]{\genfrac{(}{)}{0pt}{}{#1}{#2}}

\usepackage{cite}
\usepackage{amssymb}

\ifCLASSINFOpdf
  \usepackage{graphicx}
  \graphicspath{{./fig/}}
\else
\fi

\usepackage[cmex10]{amsmath}

\begin{document}
\title{Minimum Energy Source Coding for Asymmetric Modulation with Application to RFID}

\author{Farzad~Hessar, ~\IEEEmembership{Student Member,~IEEE}
        and~Sumit~Roy,~\IEEEmembership{Fellow,~IEEE} \\ \{farzad, sroy\}@u.washington.edu
\thanks{Contact Author: farzad@u.washington.edu}
}

\maketitle

\begin{abstract}
\boldmath
Minimum energy (ME) source coding is an effective technique for efficient communication with energy-constrained devices, such as sensor network nodes. In this paper, the principles of generalized ME source coding is developed that is broadly applicable. Two scenarios - fixed and variable length codewords - are analyzed. The application of this technique to RFID systems where ME source coding is particularly advantageous due to the asymmetric nature of data communications is demonstrated, a first to the best of our knowledge.
\end{abstract}

\begin{IEEEkeywords}
Radio Frequency Identification, Minimum Energy, Source Coding, Energy Harvesting, EPC Global.
\end{IEEEkeywords}

\IEEEpeerreviewmaketitle

\section{Introduction}

Energy constrained wireless devices increasingly require a more energy efficient wireless protocol stack to reduce power consumption and extend operation lifetime. Such energy efficiency may be obtained at multiple levels - from improved circuits, to energy efficient coding/modulation (link) and higher (multiple access and applications) layers of the protocol stack. In this work, we show how minimum energy (ME) source coding may present new opportunities for energy efficiency, as part of an overall efficient wireless stack architecture.

Radio Frequency Identification technology (RFID) is a prime example of a system with serious energy constraints at its key components. An RFID system comprises of a {\em reader} (interrogator) that initiates write/read of data to/from RFID tags; and a {\em tag} (transponder) that contains information (such as from a sensor integrated onto tag) to be transmitted to reader.  Uplink communication in RFID systems is based on {\em backscatter modulation}, a {\em passive} technique whereby a portion of the impinging RF carrier sent by the reader on the downlink is reflected by tag antenna. By modulating the tag antenna impedance, tag data may be transmitted to the reader without requiring any active transceiver components on-tag.
Common RFID tags may be fully passive, i.e. contain no battery or energy sources. Naturally, such tags must utilize some energy harvesting/scavenging mechanism (whereby it converts any impinging AC signal at it's antenna to  DC voltage) for achieving desired operational lifetime.

While backscatter uplink provides a potentially low power solution in severely energy limited components, it  must nevertheless be optimized to improve power efficiency. In this work, we explore familiar trade-offs between bandwidth and power  \cite{conf:erin99,conf:prakash03} within a framework of source coding, but with a focus on energy consumption. For example,  channel coding schemes like Turbo/LDPC codes are now commonplace in {\em high-speed} digital communications that achieve power efficiency (less power per decoded symbol required for a given error performance) close to predicted Shannon bounds,  but the decoders required are very complex. However, RFID uplink rates are considerably lower\footnote{The highest rate is $640$ Kbps on the uplink in Gen-2 EPC standard.}, suggesting that power efficiencies may be obtained by a sensible {\em joint source-channel coding} approach that explores the role of appropriate source coding, without requiring complex implementation circuitry on-tag.

Source coding involves mapping the original raw information symbols (typically, an i.i.d binary sequence) to another sequence of 1's/0's (in case of binary coding) with desired length and properties. In variable-length binary source coding, the length (number of bits) of a coded symbol is chosen based on the probability of occurrence of that symbol, so as to {\em minimize the average length}, resulting in minimum (average) length source codes. In contrast, we propose
minimum energy (ME) source coding - where the metric of interest is the {\em (average) energy} required for transmission of symbols \cite{conf:erin99,conf:prakash03,conf:wang01,conf:tang05}. The objective is to minimize the average energy required for a set of symbols, instead of the average symbol length \footnote{Note that the two are equivalent if the energy cost of transmission is independent of the symbol (0/1) being sent.}.
Hence, ME source coding is particularly beneficial to systems utilizing {\em asymmetric modulation}, i.e. with different transmission costs for bit-1 versus bit-0 as in orthogonal on-off keying (OOK). In such cases, ME source coding reduces energy at the expense of average symbol length. This loss in terms of rate reduction (or equivalently bandwidth expansion) is acceptable for applications such as RFID  or other sensornets, where the transmission rate is low - order of few kilo bits per second (Kbps) -  and transmission distances are in the range of few meters \cite{conf:wang01,conf:prakash03,conf:tang05}.

 Our primary contribution  in this work is a new and general formulation of the minimum energy source coding problem with both fixed and variable length coding techniques as two solution for this problem. It generalizes  results from previous works, notably \cite{conf:khan08,conf:kim05,conf:erin99,conf:prakash03} to backscatter modulation. Energy performance of asymmetric modulation under ME source coding is evaluated and corresponding parameters are optimized. Sec. II reviews related works on ME source coding; our ME source coding formulation is presented in Sec. III. Sec. IV provides a fixed length solution to ME source coding problem and Sec. V. formulates variable length coding solution. Simulation results are presented in Sec. VI. RFID application of ME source coding is evaluated in Sec. VII and finally VIII concludes the paper.

\section{Related Works}
Energy efficient source coding has been considered by \cite{conf:wang01,conf:erin99,conf:mohorko10,conf:khan08,conf:kim05,conf:tang05,conf:prakash03}. Wang et al. \cite{conf:wang01} presented several energy minimization techniques for a practical short range asymmetric micro-sensor system, including efficient modulation schemes, appropriate multiple access protocols, and a fast turn-on transmitter architecture. It shows that non-coherent M-FSK is more energy efficient than M-PSK/M-QAM. Khan \cite{conf:khan08} provides a source coding and modulation technique for reducing multiple access interference as well as for reducing power in an MC-CDMA network. They choose OOK for MC-CDMA and add redundant zero bits to reduce total energy consumed. Mohorko et al. \cite{conf:mohorko10} discuss network lifetime and how source coding is useful in reducing power consumption, for a IEEE 802.15.4 based monitoring of electrocardiogram signals after a two-step compressions including autoregressive predictive coding followed by Huffman entropy coding. Kim \cite{conf:kim05} proposed a modified ME source coding for DS-CDMA systems which outperforms DS-CDMA systems in both energy consumption and bit error rate.

A novel ME source coding technique was introduced by Erin \cite{conf:erin99} and extended by Prakash \cite{conf:prakash03}, for OOK with zero transmission power for bit-0. The method uses the following two steps: a) First, a set of codewords with minimum number of high bits (which results in minimum transmission power) is selected; b) Second, these codewords are assigned to symbols such that the code with lower number of bit-1 are assigned to symbols with higher probability. Table \ref{tb:ErinME} illustrates the mapping table for $k=2$ and $k=3$, where each source symbol of length $k$ is mapped to a codeword of length $n=2^k-1$. Clearly, the code rate $r=\frac{k}{2^k-1}$ is very small for large values of $k$; hence this method is highly bandwidth inefficient in general ($r=0.26 (0.031)$ for $k=4(8)$).

Our work is based on allowing variable length codewords that immediately improves upon the previous source coding schemes in terms of bandwidth efficiency for a given average energy cost. Determining examples of such improved coding techniques does not require great effort; Table \ref{tb:ImprovedErinME} presents a prefix code with lower average length than the code shown in Table \ref{tb:ErinME}, with same energy per symbol (has the same number of bit-1) and is uniquely distinguishable because of prefix property. The average length for this scheme is $L = \frac{1}{2}\left[2^k+1-2^{-(k-1)}\right]$ which is half that in Table \ref{tb:ImprovedErinME} for $k>>1$.

\begin{table}[!t]
\renewcommand{\arraystretch}{1.3}
\caption{Mapping Table for ME Source Coding by Erin\cite{conf:erin99}}
\label{tb:ErinME}
\centering
\begin{tabular}{|c|c||c|c|}
\hline
\bfseries Source bits & \bfseries Codeword & \bfseries Source bits & \bfseries Codeword \\
\hline\hline
00 & 000 & 000 & 0000000\\
01 & 001 & 001 & 0000001\\
10 & 010 & 010 & 0000010\\
11 & 100 & 011 & 0000100\\
   &     & 100 & 0001000\\
   &     & 101 & 0010000\\
   &     & 110 & 0100000\\
   &     & 111 & 1000000\\
\hline
\end{tabular}
\end{table}

\begin{table}[!t]
\renewcommand{\arraystretch}{1.3}
\caption{Improved Mapping Table for ME Source Coding}
\label{tb:ImprovedErinME}
\centering
\begin{tabular}{|c|c||c|c|}
\hline
\bfseries Source bits & \bfseries Codeword & \bfseries Source bits & \bfseries Codeword \\
\hline\hline
00 & 000 & 000 & 0000000\\
01 & 1   & 001 &       1\\
10 & 01  & 010 &      01\\
11 & 001 & 011 &     001\\
   &     & 100 &    0001\\
   &     & 101 &   00001\\
   &     & 110 &  000001\\
   &     & 111 & 0000001\\
\hline
\end{tabular}
\end{table}

Prior work on ME source coding for OOK modulation has typically assumed that the transmitter uses zero power for bit-0 and non-zero power for bit-1. Therefore, the obvious choice would be to increase number of bit-0 and decrease bit-1 to achieve a more energy efficient system. In this paper, we consider systems based on backscatter modulation as in RFID applications, whereby modulation consumes zero power for transmitting both bit-1 and bit-0. Therefore, the previous notion of energy consumption for bit-0/bit-1 that is solely based on modulation energy is no longer applicable. A more general definition of transmission cost is needed which must consider energy harvesting, in case of RFID context, as part of the cost metric. As a result of this difference, ME source coding approaches introduced in \cite{conf:mohorko10,conf:khan08,conf:kim05,conf:tang05,conf:prakash03} are not only inapplicable for RFID systems but they may also increase average energy as discussed in following sections.

\section{Problem Formulation}

Let's assume a binary transmitter with following parameters:
\begin{itemize}
	\item $T_0$/$T_1$: Transmission time for bit-0/bit-1 (seconds) equal to the duration of the waveform representing bit-0/bit-1.
	\item $\beta_0$/$\beta_1$: Transmission cost for sending bit-0/bit-1
	\item $A=\{a_1,a_2,...,a_M\}$: Set of all symbols generated by the source with probabilities $P=\{p_1,p_2,...,p_M\}$ where  $\sum_{i=1}^{M}{p_i}=1$ and we order $p_1\leq p_2 \leq ... \leq p_M$ w.l.o.g.
	\item $C=\{c_1,c_2,...,c_M\}$: Set of all codewords assigned to symbols in A via 1:1 mapping  $a_i \rightarrow c_i$.  The length of each codeword (in terms of number of 0/1) is represented by $l_i$.
\end{itemize}

The data source generates $M$ different symbols with the average length of $L_{src}=H(P)$ where $H(.)$ is the usual entropy function for a discrete source. Hence:
\begin{align}
L_{src} = -\sum_{i=1}^{M}{p_i\log_2(p_i)}
\end{align}
The average duration of a source symbol is $T_{src}=\frac{T_0+T_1}{2}L_{src}$, and the resulting rate of transmission (symbol/sec) by sending uncoded source symbols is:
\begin{align}
R_{src} \,= \; \frac{1}{T_{src}}=\frac{2L_{src}}{T_0+T_1}
\end{align}

Every symbol $a_i$ is mapped to a codeword $c_i$ that involves $N_0(c_i)$ number of zeros and $N_1(c_i)$ number of ones. The average length of codewords $L_{code}$ in terms of number of bits is:
\begin{align}
L_{code} = \frac{1}{M}\sum_{i=1}^{M}{p_il_i}
\end{align}
and the average duration of codewords is:
\begin{align}
T_{code} = \sum_{i=1}^{M}{p_i\left[ N_0(c_i)T_0 + N_1(c_i)T_1 \right]}
\end{align}
which results in coded transmission rate as $R_{code}\, = \,\frac{1}{T_{code}}$. The rate reduction factor (or equivalently bandwidth expansion factor) $\eta$ is:
\begin{align}
\label{eq:ratefact}
\eta &= \frac{R_{src}}{R_{code}}\nonumber \\
     &= \frac{2L_{src} \sum_{i=1}^{M}{p_i\left[ N_0(c_i)T_0 + N_1(c_i)T_1 \right]} }{ T_0+T_1 }
\end{align}

The main metric of interest is average energy cost of transmission in the coded system which is defined as:
\begin{align}
\label{eq:txcost}
\overline{\beta_{code}}=\frac{1}{M}\sum_{i=1}^{M}{p_i\left(\beta_0N_0(c_i)+\beta_1N_1(c_i)\right)}
\end{align}

The ME source coding problem is defined in terms of average transmission cost (\ref{eq:txcost}) and rate reduction factor (\ref{eq:ratefact}) as \\

{\em Problem (Optimal ME Source Coding):}
Find a set of uniquely distinguishable codewords $C=\{c_1,...,c_M\}$ such that mapping source symbols $A=\{a_1,...,a_M\}\rightarrow C$ results in minimum transmission cost, i.e.,
\begin{align}
\label{eq:definition}
\overline{\beta_{opt}}= \underset{\{c_1,...,c_n\}}{\text{minimize}}& \left[ \frac{1}{M}\sum_{i=1}^{M}{p_i\left(\beta_0N_0(c_i)+\beta_1N_1(c_i)\right)} \right] \nonumber \\
\mbox{s.t.} \,\,\,\,\,\, &\frac{R_{src}}{R_{code}} \leq \eta_{thr}
\end{align}
where $\eta_{thr}$ is the maximum acceptable rate reduction factor in the application.

Depending on the type of application, various special cases of the problem may be solved with a unique method. Some of these  include:
\begin{itemize}
	\item $T_0=T_1$: In many wireless applications, the waveforms that are used for transmission of bit-1 and bit-0 are of the same duration(although this is not the case for RFID standards). This simplifies (\ref{eq:ratefact}) to be independent of $T_0$ and $T_1$. Thus, regardless of how we assign 0/1 to different symbols, decreasing the average length of the code words guarantees higher throughput.
	
	\item $\beta_0 = \beta_1$: For the case of symmetric modulation where bit-1 and bit-0 are equal cost, (\ref{eq:txcost}) reduces to minimizing average length of the codewords, the same as regular source coding techniques.
	
	\item The codewords $C=\{c_1, c_2, ... , c_M\}$ can be fixed or variable length. The optimization problem is significantly different in these two cases.
\end{itemize}
The next section is focused toward solving the problem for the case of fixed length codewords.

\section{Minimum Energy Coding for Fixed Length Codewords}
Let's assume each codeword in $C=\{c_1,c_2,...,c_M\}$ is of length $n$. There are $2^n$ number of codewords of length $n$ and obviously $M\leq 2^n$. Each codeword $c_i$ is composed of $n_{i,0}$ number of zeros and $n_{i,1}$ number of ones; $n_{i,0}+n_{i,1}=n$. The cost of transmitting codeword $c_i$ is $f(c_i)=\beta_1n_{i,1} + \beta_0n_{i,0}$.

Without loss of generality, let's assume  $\beta_0<\beta_1 \, , \Delta\beta=\beta_1-\beta_0$. Therefore:
\begin{equation}
f(c_i)=n\beta_0+\Delta\beta n_{i,1}
\label{eq:codecost}
\end{equation}
This equation clearly shows that minimizing the number of bit-1, $n_{i,1}$, does not necessarily decrease cost function unless $\beta_0=0$.

The average cost for the set of M symbols is:
\begin{align}
\overline{\beta_{code}} &= \sum_{i=1}^{M}{p_if(c_i)} =  \sum_{i=1}^{M}{p_i\,n\,\beta_0+p_i\,\Delta\beta\,n_{i,1}} \nonumber \\
								 &= n\,\beta_0 + \Delta\beta\sum_{i=1}^M{p_i\,n_{i,1}}
\label{eq:OptProb}
\end{align}

Let's sort the codewords in terms of number of bit-1's from 0 to $n$ as $c_0\,\leq c_1\,\leq c_2\,\leq ... \leq\,c_{2^n-1}$. Obviously, the best coding scheme selects the codewords with minimum number of bit-1's as $C=\{c_0\,\leq c_1\,\leq c_2\,\leq ... \leq\,c_{M-1}\}$. For a code of fixed-length $n$, there are $\binom{n}{k}$ codewords with k number of 1's. Assume $l_{\min}(n)$ is defined as:
\begin{equation}
l_{\min}(n) \triangleq \min l \,\,\,\,\, \mbox{s.t.    }   \sum_{k=0}^{l}{\binom{n}{k}}\,\geq M
\end{equation}

The average cost for the fixed length of $n$ and assuming uniform distribution is:
\begin{align}
\overline{\beta_{code}} &= n\beta_0 + \nonumber \\
&+  \frac{\Delta\beta}{M} \left[\sum_{i=0}^{l_{\min}-1}{i\,\binom{n}{i}} + l_{\min}\left(M-\sum_{k=0}^{l_{\min}-1}{\binom{n}{k}}\right)\right]
\label{eq:fixedLenCost}
\end{align}
Note the distribution is assumed uniform as a general solution for a source with unknown statistics.

The cost function in (\ref{eq:fixedLenCost}) depends only on $n$. Therefore, the optimization problem in the case of fixed length coding is simplified to finding the best $n$. Note that we relieved the bandwidth constraint in (\ref{eq:definition}) for the fixed code length case, so as to achieve the best energy optimization which is acceptable in RFID applications.

\begin{align}
n_{opt} = \underset{n}{\text{ArgMin}}\left( n + \frac{(\beta_1/\beta_0)-1}{M}\left[\sum_{i=0}^{l_{\min}-1}{i\,\binom{n}{i}} + \right. \right. \nonumber \\
 \left. \left. + l_{\min}\left(M-\sum_{k=0}^{l_{\min}-1}{\binom{n}{k}}\right)\right]\right)
\label{eq:opt}
\end{align}

The optimum code length is a function of two parameters, $n_{opt}=n(M, \frac{\beta_0}{\beta_1})$. An analytical description for $n_{opt}$ needs suitable approximation for the sum of binomial coefficients, for which the usual candidates are `Poisson' and `Gaussian' approximations. Our results reveal that these are not appropriate because $n_{opt}$ depends largely on the binomial coefficients with small values of $k$ relative to $n$ in $\binom{n}{k}$. These correspond to estimating the tail of Gaussian/Poisson density which is prone to maximum error.

\section{Minimum Energy Coding For Variable Length Codewords}
Design of a variable-length code book is significantly different from fixed-length code book in the sense that it must be {\em uniquely decodable}, i.e. for every sequence of code words $C_{i_1}C_{i_2}C_{i_3}...C_{i_n}$ there must be a one-to-one mapping to a sequence of source symbols $a_{i_1}a_{i_2}a_{i_3}...a_{i_n}$ such that the code words are correctly decoded at the destination. Therefore, finding ME code in this case is not accomplished solely by selecting symbols with minimum number of bit-1 and requires further constraints to guarantee decodability.

The condition for {\em uniquely decodable} codes is introduced in \cite{bk:IT_Cover} which is based on non-singularity of all extensions of codewords (all possible sequences). Note this extension (sequence) is unlimited in length and is difficult to embed as a constraint in the optimization problem for ME source coding. One solution is to use {\em instantaneous} or {\em prefix} codes, whereby no codeword is a prefix of any other codeword. These type of codes are proven to be uniquely decodable \cite{bk:IT_Cover}. Furthermore, they support fast decoding because receiver does not have to wait until the entire sequence is received to begin decoding.

\begin{figure}[t]%
\centering
\includegraphics[width=\figwidth]{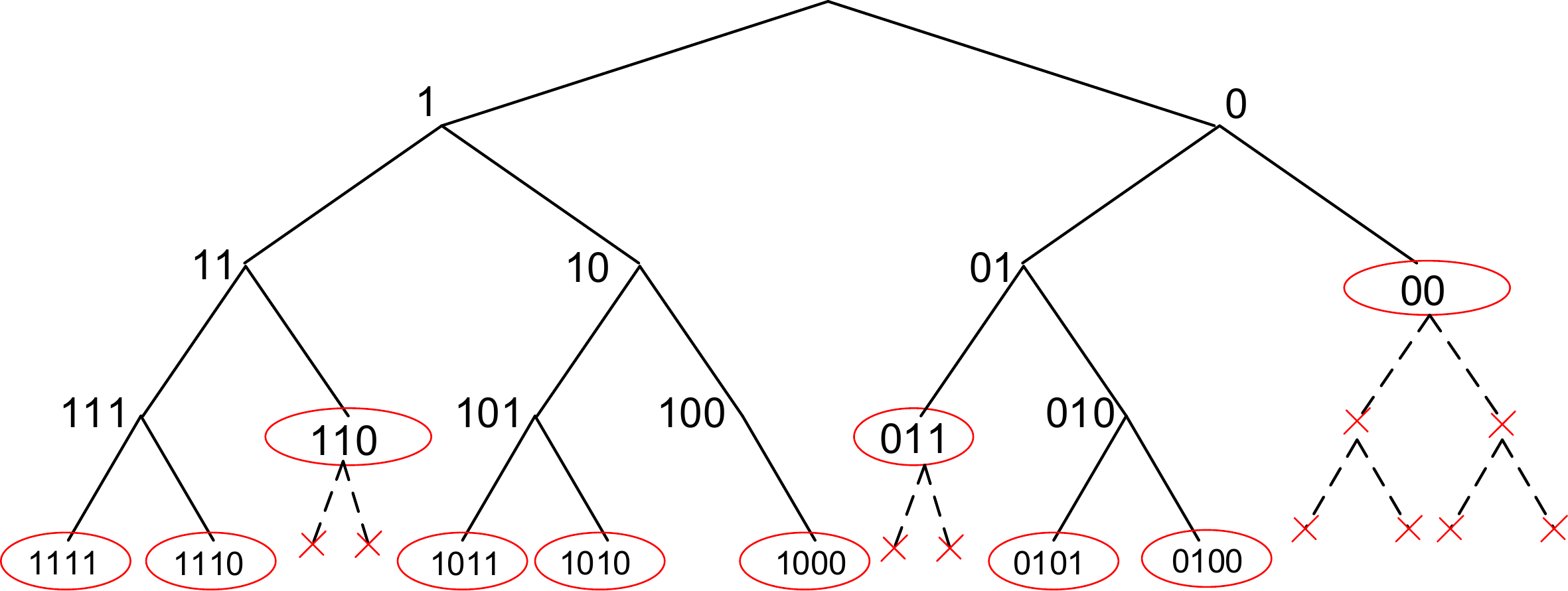}%
\caption{Prefix codes can be generated using binary trees and selecting leaves as the codewords. Dashed lines represents those codewords that cannot be selected in the codebook because their parent (their prefix) is already selected.}%
\label{fig:tree}%
\end{figure}

Prefix codes can be generated on binary trees by selecting tree leaves for codewords (no leaf is a prefix of another leaf) as shown in Fig. \ref{fig:tree}. Consider a binary tree of depth $dp$. The longest codeword generated by this tree is $dp$ bits long and the total number of codewords contained in the tree is $Q=2^{dp+1}-2$. Let $\mathcal{C}=\{c_1, c_2, ..., c_Q\}$ be the ordered set of all codewords in the tree and $\mathcal{A}^{Q\times1}$ be a binary selection vector such that $\mathcal{A}(i)=1$ if $c_i$ is selected as a codeword in C and $\mathcal{A}(i)=0$ otherwise. The cost function for each codeword $c_i$, as defined before, is $f(c_i)$ and the cost vector $\mathcal{F}=[f(c_1), f(c_2), ..., f(c_Q)]^T$. The problem is now reduced to finding the selection vector $\mathcal{A}$ subject to following constraints:
\begin{itemize}
	\item Since there are M source symbols, there must be M codewords selected from $\mathcal{C}$. Therefore, $\mathcal{A}^T\mbox{{\bf 1}}^{Q\times 1}=M$ where $\mbox{{\bf 1}}^{Q\times 1}$ is a column vector of ones.
	\item Prefix condition requires that if a codeword $c_i\in\mathcal{C}$ is selected ($\mathcal{A}(i)=1$) then all codewords corresponding to children of this node in the tree cannot be selected, as shown in Fig. \ref{fig:tree}. Therefore, it can be described as $\mathcal{A}(i) + \mathcal{A}(j) \leq 1$ for all $(i,j)$ subject to node $i$ is parent of node $j$. Let's assume that codewords in $\mathcal{C}$ are numbered in a depth-first order	(left branch of each node is numbered before the right branch) then for a node at depth $d$, the next $2^{tr-d}-1$ nodes are children. Let's define a binary parent-child relationship matrix $\mathcal{P}(dp)$ as  below (for the case of $dp=2$):
\begin{align}
\mathcal{P}(dp=2) =
\left[
	\begin{array}{cccccc}
		1 & 1 & 0 & 0 & 0 & 0 \\
		1 & 0 & 1 & 0 & 0 & 0 \\
		0 & 0 & 0 & 1 & 1 & 0 \\
		0 & 0 & 0 & 1 & 0 & 1 \\
	\end{array}
\right]
\end{align}
where each row defines a parent-child pair. Due to its symmetry, this matrix can be recursively created as follows. For given $\mathcal{P}^{r\times c}(n)$
\begin{align}
\mathcal{P}(dp=n+1) =
\left[
	\begin{array}{cc}
		\mathcal{Q} & \mbox{{\bf 0}} \\
		\mbox{{\bf 0}} & \mathcal{Q}
	\end{array}
\right]	
\end{align}
where
\begin{align}
\mathcal{Q} =
\left[
	\begin{array}{cc}
		\mbox{{\bf 1}}^{q\times 1} & I(q) \\
		\mbox{{\bf 0}}^{r\times 1} & \mathcal{P}(n) \\
		\mbox{{\bf 0}}^{r\times (\frac{q}{2}+1)} & \mathcal{P}(n)
	\end{array}
\right]
\end{align}
with $q=2^{n+1}-2$ and $I(q)$ is $q\times q$ identity matrix. The prefix condition can be described as below:
\begin{align}
\mathcal{P}(dp)\mathcal{A} \leq \mbox{{\bf 1}}
\end{align}
\end{itemize}
The optimum variable length ME source code is found by solving following Integer Program:
\begin{align}
\min_{\mathcal{A}} & \,\,\, \mathcal{F}^T \mathcal{A} \nonumber \\
\mbox{ s.t.  } &\mathcal{A} \in Z^{Q} \nonumber \\
& \mbox{{\bf 1}}^{1\times Q} \mathcal{A} = M\nonumber \\
& \mathcal{P}(dp)\mathcal{A} \leq \mbox{{\bf 1}}
\end{align}
The tree depth parameter, $dp$, controls how deep the optimization algorithm traverses down the tree to find the optimum set. Ideally $dp=M-1$ to cover all possible codewords for the case $\gamma=\frac{\beta_1}{\beta_0}\rightarrow \infty$. However, for $\gamma<\infty$, $dp$ can be set much smaller than $M-1$ to speed up optimization process.

\section{Simulation Results}
In this section, simulation results are provided for both fixed and variable length source coding techniques, introduced before. This includes transmission cost function $\overline{\beta_{code}}$, $n_{opt}$ and resulting energy saving factor as a function of basic parameters $M$ and $\frac{\beta_1}{\beta_0}$, $dp$ as well as comparison of this two methods.

\subsection{Simulation Results of Fixed Length Cost Function}
\begin{figure}[t]%
\centering
\includegraphics[width=\figwidth]{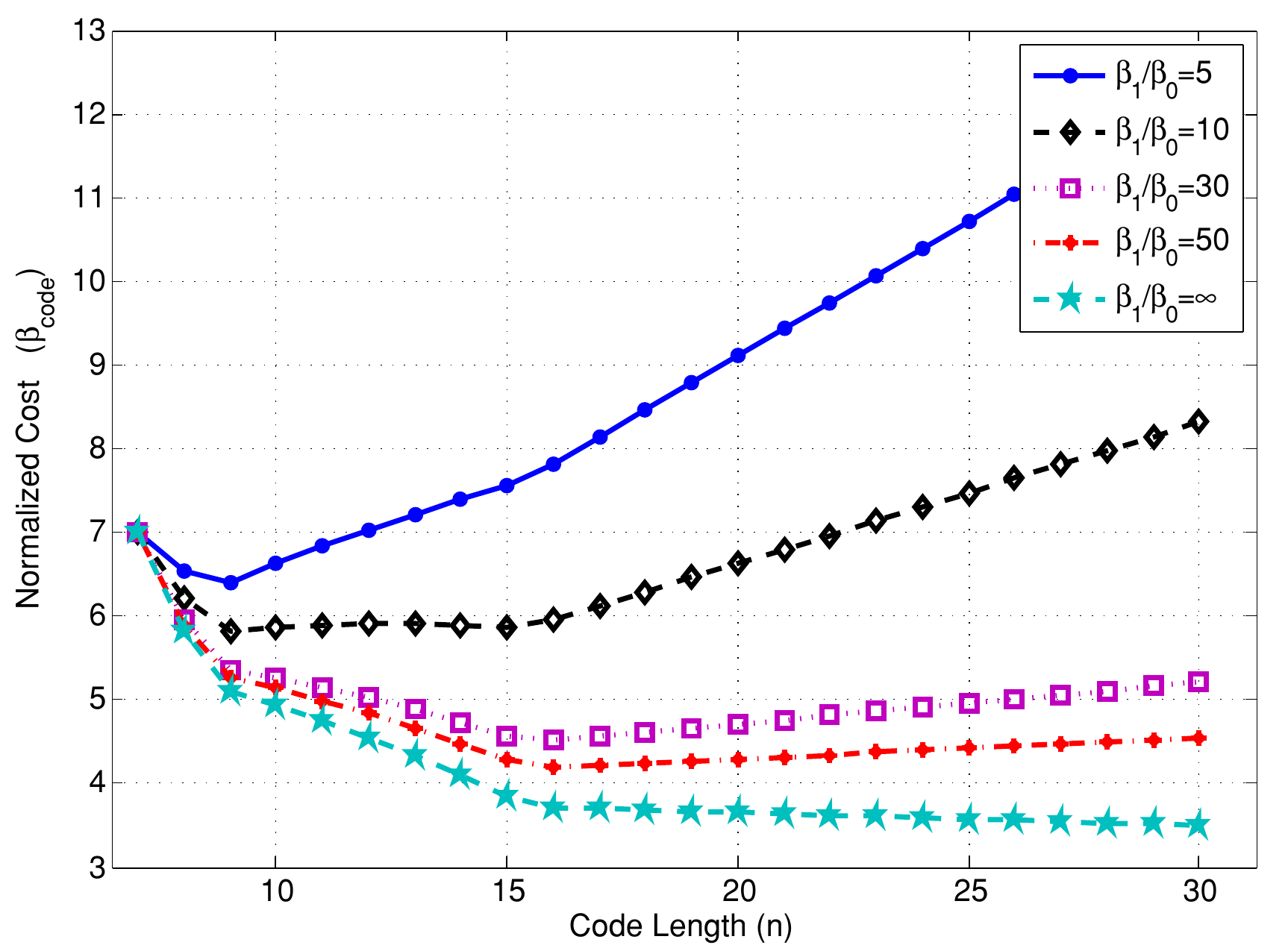}%
\caption{Average transmission cost $\overline{\beta_{code}}$ (\ref{eq:fixedLenCost}) vs. code length (n) for various values of $\gamma=\frac{\beta_1}{\beta_0}$}%
\label{fig:fxdCodeLen1}%
\end{figure}

For a case of M=128 (7-bit source symbols), Fig. \ref{fig:fxdCodeLen1} plots normalized average transmission cost $\overline{\beta_{code}}$ as a function of $n$ for various values of $\gamma=\beta_1/\beta_0$. It is inferred from this figure that for small values of $\gamma$ the optimum code length $n_{opt}$ is close to $\log_2(M)=7$. This is because cost of sending bit-0 is not significantly lower than bit-1 ($\gamma$ is close to 1) and therefore it is not cost-effective to replace a bit-1 with multiple bit-0. As $\gamma$ grows larger, the optimum code length increases up to a maximum of $n_{max}=M-1$. This maximum code length is optimum only if $\gamma=\frac{\beta_1}{\beta_0}\rightarrow \infty$, where every codeword has only one bit-1 and $M-2$ number of bit-0.

It is also understood from Fig. \ref{fig:fxdCodeLen1} that for limited values of $\gamma$, choosing $n=M-1$ is not the optimum point for average cost and it may even increase the cost beyond that of the uncoded ($n=\log_2(M)$) case. This is clearly different from the results in \cite{conf:prakash03,conf:tang05} where $n=M-1$ is considered as the general solution for all cases.

\subsection{Optimum Code Length for Fixed Length Coding}
\begin{figure}[t]%
\centering
\includegraphics[width=\figwidth]{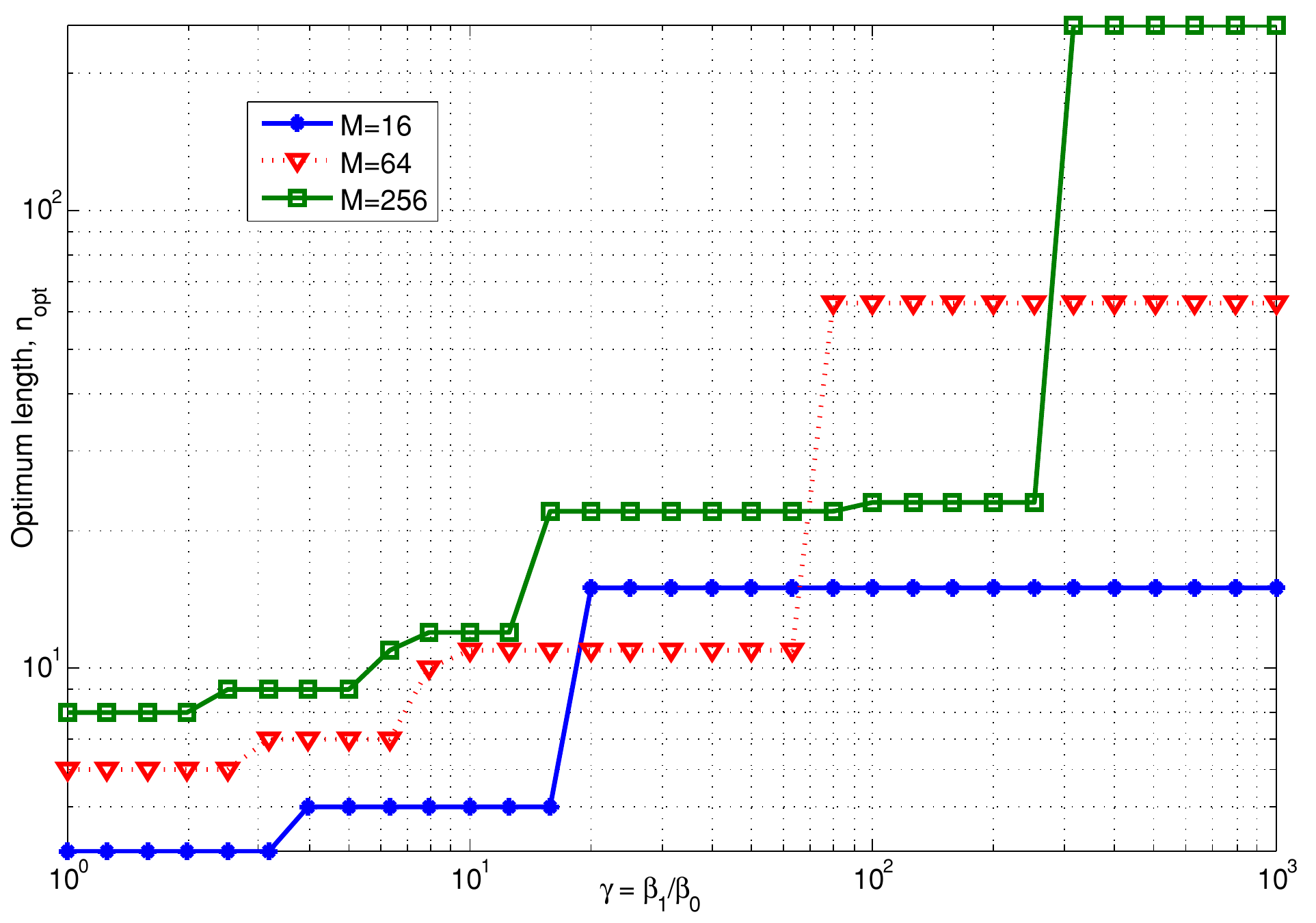}%
\caption{Optimum length $n_{opt}$ (\ref{eq:opt}) vs $\gamma$ for various values of $M$}%
\label{fig:OptNvsGamma}%
\end{figure}

The objective function in (\ref{eq:opt}) is not in a closed mathematical form, therefore the optimum length $n_{opt}$ cannot be determined analytically. Fig. \ref{fig:OptNvsGamma} plots $n_{opt}$ as a function of $M$ and $\gamma=\frac{\beta_1}{\beta_0}$. The function $n(M,\gamma)$ has following properties:
\begin{align*}
n(M, 1) = \log_2(M)\\
\lim_{\gamma\to\infty}n(M, \gamma)=M-1
\end{align*}
The code length $n(M,\gamma)$ in Fig. \ref{fig:OptNvsGamma} is a monotonically increasing function of $\gamma$ which starts from  $\log_2(M)$ for $\gamma=1$ and increases up to $M-1$ for $\gamma\rightarrow\infty$.

\begin{figure}[t]%
\centering
\includegraphics[width=\figwidth]{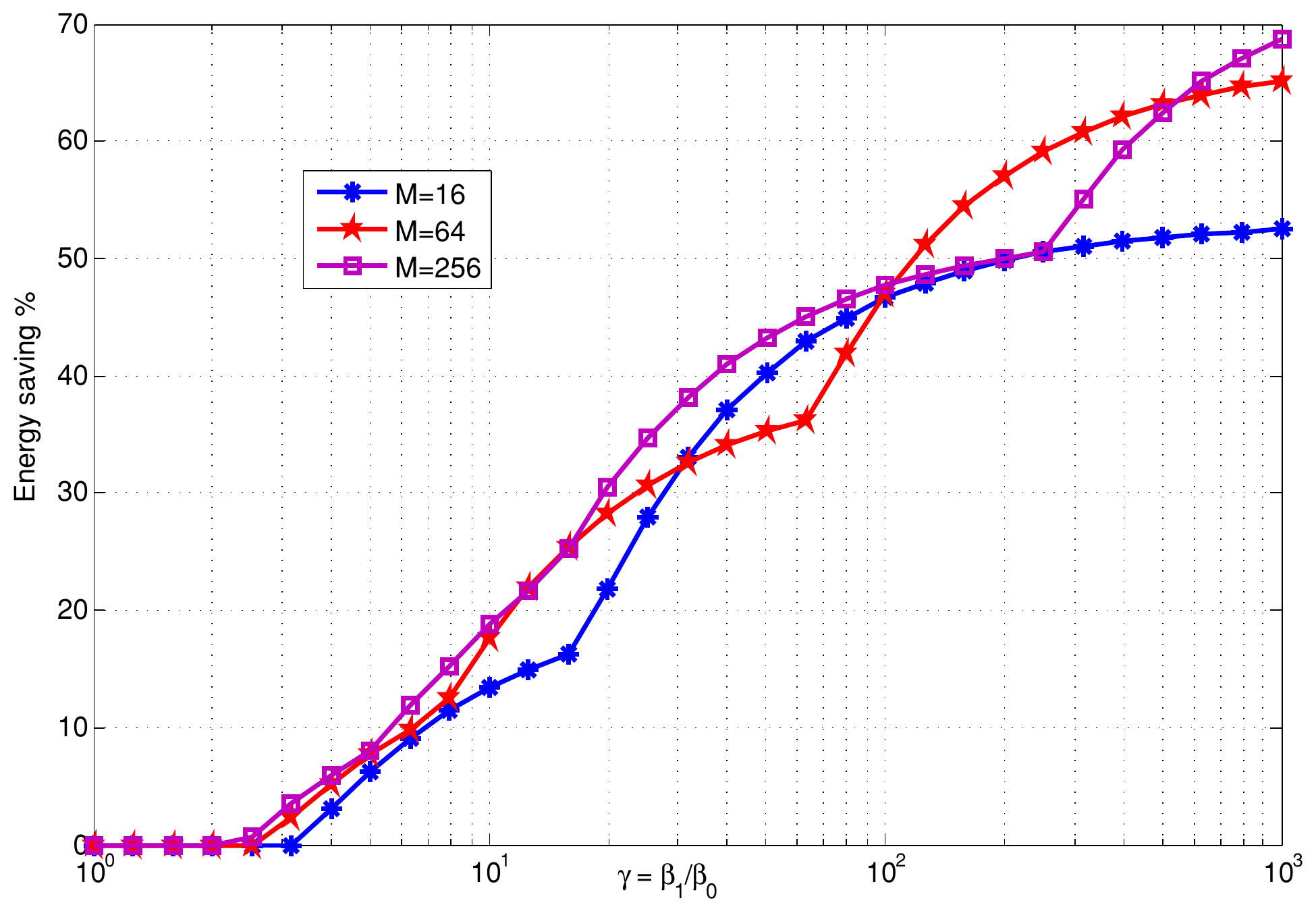}%
\caption{Energy saving percentile versus $\gamma$ for various values of $M$, fixed length coding}%
\label{fig:CostvsGamma}
\end{figure}

It is important to see how much energy is saved using this scheme compared to the original non-coded system. The average cost for the source symbols is $\beta_{src}=\frac{1}{2}(\beta_0+\beta_1)\log_2(M)$. Hence, the energy saving factor is:
\begin{align}
\epsilon&=1-\frac{2\beta_{opt}}{(\beta_0+\beta_1)\log_2(M)} \nonumber\\
&=1-\frac{2\beta_{opt}/\beta_0}{(1+\gamma)\log_2(M)}\nonumber\\
&=\epsilon(M, \gamma)
\label{eq:eSaving}
\end{align}
which is a function of cost ratio $\gamma$ and not $\beta_0$ or $\beta_1$ individually. Fig. \ref{fig:CostvsGamma} presents saved energy percentage using fixed length ME source coding as a function of $\gamma$ and for various values of $M$. More energy is saved as the cost ratio increases which is because of increased cost distance of bit-0 and bit-1. Therefore, the maximum energy saving is achieved for $\gamma\rightarrow\infty$ which is limited to:
\begin{align}
\epsilon_{\max} &= \lim_{\gamma\rightarrow\infty}{\epsilon(M,\gamma)}\nonumber\\
&= \lim_{\gamma\rightarrow\infty}{1-\frac{(M-1)\beta_1+(M-1)^2\beta_0}{M/2(\beta_0+\beta_1)\log_2(M)}}\nonumber\\
&=\lim_{\gamma\rightarrow\infty}{1-\frac{M-1}{M}\frac{2(\gamma+M-1)}{(\gamma+1)\log_2(M)}}\nonumber\\
&=1-\frac{2(M-1)}{M\log_2(M)} \nonumber\\
&\approx 1-\frac{2}{\log_2(M)} \,\, \mbox{ for } M>4
\end{align}

\subsection{Variable Length Coding Performance}
\begin{figure}[t]%
\centering
\includegraphics[width=\figwidth]{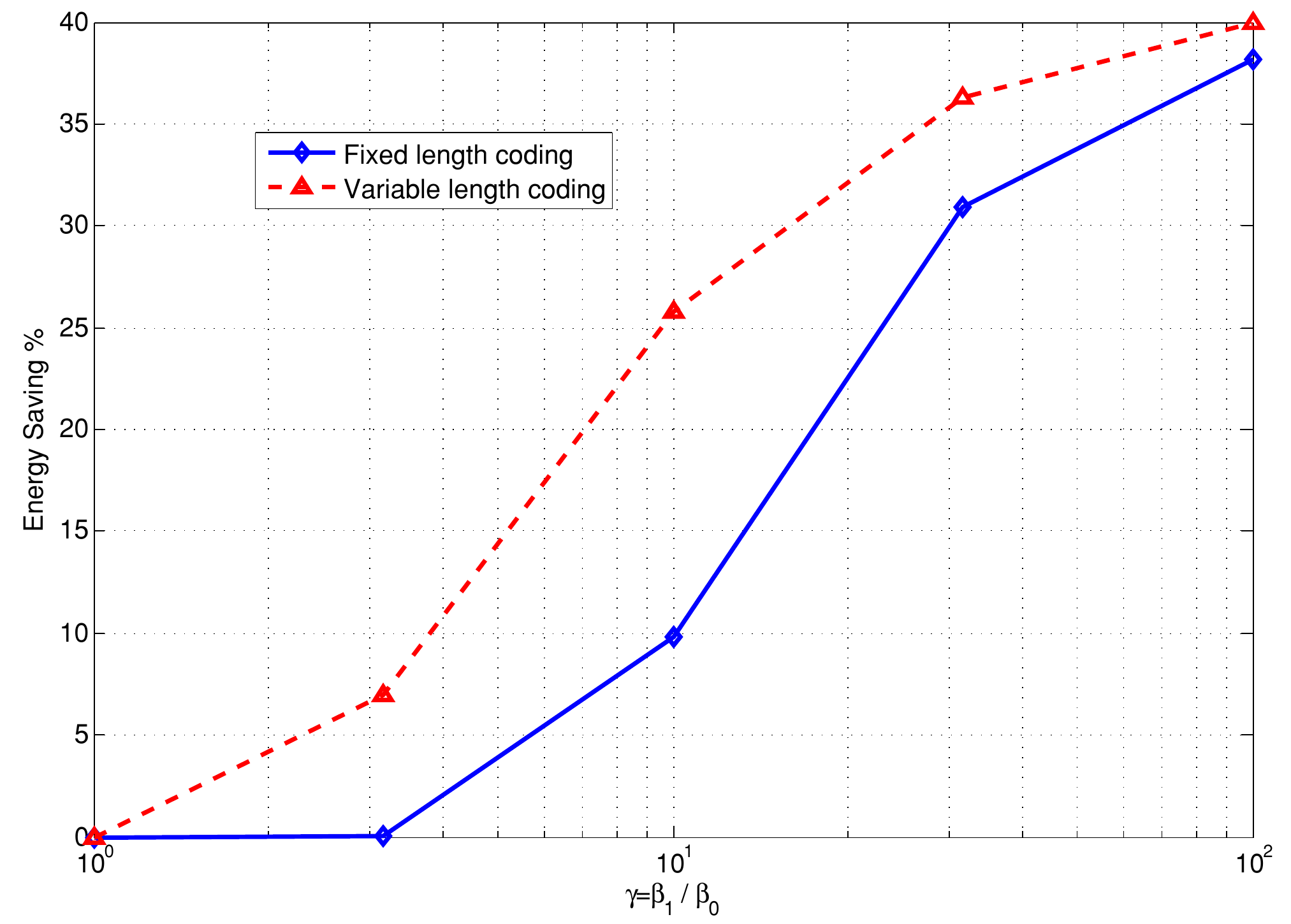}
\caption{Energy saving percentile versus $\gamma$ for variable length coding versus fixed length coding, $M=8$}
\label{fig:SavingVariable}
\end{figure}

Performance of variable length coding is a function of $M$ and $\gamma$ in a similar manner to fixed length coding. The best energy saving (relatively) is achieved for large values of $\gamma$ for which the resulting codebook has a similar structure to codes in Table \ref{tb:ImprovedErinME}. Therefore, for $\gamma\rightarrow\infty$:
\begin{align}
\epsilon_{\max} &= \lim_{\gamma\rightarrow\infty}{\epsilon(M,\gamma)}\nonumber\\
&= \lim_{\gamma\rightarrow\infty}{1-\frac{2\gamma(M-1)+M(M-1)}{M(1+\gamma)\log_2(M)}}\nonumber\\
&\approx 1-\frac{2}{\log_2(M)} \, \mbox{ for } M>4
\end{align}
Clearly, for very large values of $\gamma$, variable length coding has the same energy saving performance as fixed length coding. On the other hand, for small and moderate values ($2\leq\gamma\leq 100$), there is a significant improvement in using variable-length versus fixed-length codewords as shown in Fig. \ref{fig:SavingVariable}. 
An additional energy saving factor of up to 15\% is achievable through variable-length coding for $M=8$. 
\begin{table}[!t]
\renewcommand{\arraystretch}{1.3}
\caption{Mapping table for fixed and variable length coding, $\gamma=5$, $M=8$}
\label{tb:VarLength}
\centering
\begin{tabular}{|c|c||c|}
\hline
\bfseries Source bits & \bfseries Fixed length code & \bfseries Variable length code \\
\hline\hline
000 & 000 & 11\\
001 & 001 & 10\\
010 & 010 & 01\\
011 & 011 & 001\\
100 & 100 & 0001\\
101 & 101 & 00001\\
110 & 110 & 000001\\
111 & 111 & 000000\\
\hline
\hline
Average Cost & 9 & 7.75 \\
\hline
\end{tabular}
\end{table}

Table \ref{tb:VarLength} shows an example of ME source coding for both fixed and variable codes, $M=8$ and $\gamma=5$. This is a sample case where fixed length coding does not improve energy efficiency and the resulting codebook is essentially the same as the source bits. Variable length coding, however, has reduced energy consumption by 14\%. It is noteworthy that variable length codebook highly depends on $dp$, the code tree depth. It controls how deeply the algorithm searches through the prefix codebook tree to find the optimum set. Clearly, decreasing $dp$ will speed up the process but in order to guarantee this method will find the best results, $dp>=M-1$.

\section{Cost Ratio in Passive RFID Systems}
As the results of previous sections suggest, most of the analysis as well as simulation results are function of cost ratio $\gamma=\frac{\beta_1}{\beta_0}$. In this section, $\gamma$ is evaluated for passive RFID systems.

The physical layer in passive/semi-passive RFID systems is based on signal backscattering where transmitter does not transmit its own power but it modulates backscattered signal. Since RFID transmitter (tag) does not send its own power, definition of bit-1/bit-0 cost is very much different from regular wireless systems. Furthermore, the incident waves are used as an input source of energy through energy scavenging process. Therefore, a reasonable definition for energy cost in RFID systems is the difference between consumed energy for processing and harvested energy.

\begin{figure}[t]%
\centering
\includegraphics[width=\figwidth]{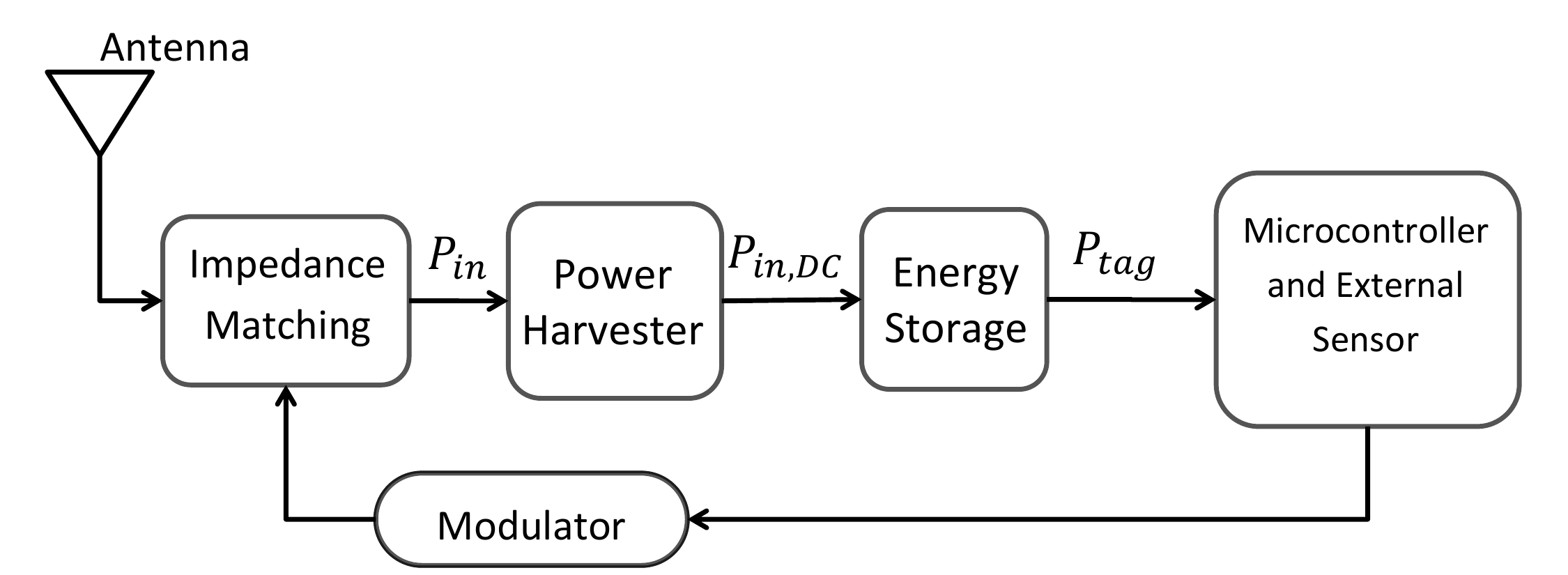}%
\caption{Block diagram for a full/semi-passive RFID tag}%
\label{fig:Blocks}
\end{figure}

Fig. \ref{fig:Blocks} presents the block diagram for a passive/semi-passive RFID tag \cite{Thesis:Alanson}. The antenna front-end is directly connected to an impedance matching circuit that is controlled by backscattering modulator. This block controls the input impedance between two nominal values of $Z_{in}=Z^*_{ant}$ and $Z_{in}=0$ which results in maximum and zero power absorption (zero or maximum reflected power), respectively. We assume the former is used for sending bit-0 and the latter for bit-1.

The absorbed (harvested) power is rectified through {\em Power Harvester} block and is stored in {\em Energy Storage} block (usually capacitor or rechargeable battery). The bit-1/bit-0 modulation cost, defined in previous sections, is stated in terms of the difference between effective DC power delivered to energy storage and energy consumed by power harvester. The modulation cost is therefore defined as:
\begin{itemize}
	\item $\beta_0=\left[(P_{tag}-P_{in,DC})T_0\right]^+$
	\item $\beta_1=P_{tag}T_1$
\end{itemize}
where $P_{tag}$ is power consumption of the tag circuit (processing power) and $P_{in,DC}$ is rectified DC power (harvested power), as shown in Fig. \ref{fig:Blocks}. The `+' term in definition of $\beta_0$ stands for positive part of the cost ($[.]^+=\max(., 0)$ ) and it is used to have a bounded optimum results. Otherwise if $P_{tag}\,<\,P_{in,DC}\rightarrow\beta_0\,<\,0$ and therefore we could add unlimited number of bit-0 and the average cost $\overline{\beta_{code}}$ keeps decreasing with no lower bound.

The cost ratio should be described in terms of transmitted carrier by {\em reader} and RFID tag specification. Let's $P_{T}$ be the carrier power and $P_{in}$ the input power at the output of impedance matching block:
\begin{align}
P_{in} = P_T\left(\frac{\lambda}{4\pi r}\right)^2G_TG_RL_P
\label{eq:Friis}
\end{align}
where $G_T/G_R$ are transmitter/receiver antenna gains and $L_P$ is polarization loss \cite{paper:AlansonWISP}. The rectified voltage at the output of harvester block is \cite{bk:Dobkin}:
\begin{align}
V_{DC} = 2N(V_{ant}-V_t) \\
V_{ant}=2\sqrt{2R_{ant}P_{in}}
\label{eq:Voltage}
\end{align}
with the inherent assumption of perfect matching circuit at the antenna front-end; otherwise there should be another factor counting for power-loss due to mismatch. The factor $N$ is number of voltage multiplication stages and $V_t$ is the threshold voltage for rectifying diodes.

The effective DC power delivered to energy storage could be related to actual input power $P_{in}$ through an efficiency factor $\eta_{rect}=\frac{V_{DC}}{V_{DC}+2NV_t}$ \cite{bk:Dobkin}. Thus:
\begin{align}
P_{in,DC} = \left[1-\frac{V_t}{2\sqrt{2R_{ant}P_{in}}}\right]P_{in}
\label{eq:Pindc}
\end{align}
and the cost ratio $\gamma$ is described as:
\begin{align}
\gamma = \frac{\beta_1}{\beta_0} = \frac{P_{tag}T_1}{\max{\left[0, (P_{tag}-P_{in,DC})T_0\right]}}
\label{eq:cstrtioRFID}
\end{align}
Two cases are possible for $\gamma$ based on how rectified power compares with $P_{tag}$:
\begin{enumerate}
	\item $P_{tag}\,>\,P_{in,DC}$: In this case, rectified input power is less than enough to run the circuit even though perfect impedance matching is performed and maximum power is absorbed. Therefore, careful choice of code length $n$ (fixed length coding) or codeword set (variable length coding) should be taken otherwise it could make the source coding less efficient than uncoded system, as shown in Fig. \ref{fig:fxdCodeLen1}.
	\item $P_{tag}\,<\,P_{in,DC}$: In this case $\beta_0=0$ and sending bit-0 is essentially of no cost to transmitter. The cost ratio factor $\gamma\rightarrow\infty$ and the optimum code length in this case is therefore $n=M-1$ which results in $\epsilon_{\max}\approx1-\frac{2}{\log_2(M)}$.
	
\end{enumerate}

\subsection{Comparison to current standard}
The current standard for RFID, EPC Gen2 Class 1 provides two types of line coding, namely FM0 and Miller. Both FM0 and Miller are composed of full-wave high/low pulses as well as half-wave pulses. The Miller encoding is eventually multiplied by a sub-carrier that turns all pulses to half-wave. An interesting question is raised in comparison of modulation by full-wave pulses and half-wave pulses in terms of the overall energy consumption.

\begin{figure}[t]%
\centering
\includegraphics[width=\figwidth]{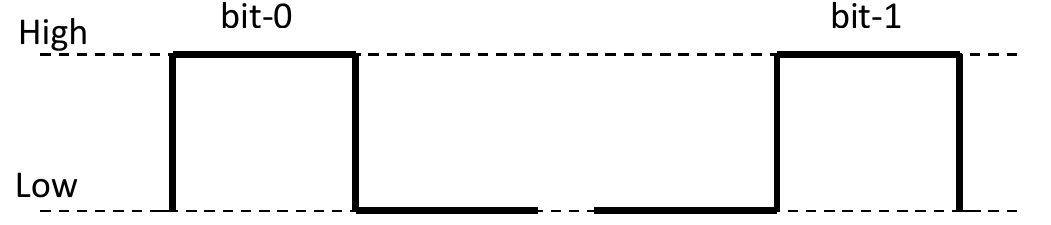}
\caption{Half wave pulse shaping in Miller and FM0 encoding}%
\label{fig:miller}%
\end{figure}

Full wave modulation leads to the same results provided in previous sections. But in half-wave modulation, the bit-0 is represented by a pulse of half-low half-high in amplitude and bit-1 is modulated by the same pulse with a reverse ordering, shown in Fig. \ref{fig:miller}.

Because of symmetry, the cost of sending bit-0 and bit-1 are the same in this type of line coding and the average cost is: $\beta_{uncoded}=\log_2(M)(\beta_0+\beta_1)/2$. This cost is the same as uncoded full wave modulation and therefore migrating from this symmetric coding to non-symmetric modulation will provide the same energy saving as in Fig. \ref{fig:CostvsGamma} and Fig. \ref{fig:SavingVariable}.

\section{Conclusion}
We introduced a generalized definition for ME source coding problem as an effective technique for energy reduction in wireless low power systems. This generalization extends application of this method beyond OOK modulation to every non-symmetric modulation system. The basic trade-off is to compromise bandwidth for energy in a practical, non-complex procedure that can be implemented in sensor network nodes or RFID tags.

There could be various solutions to the introduced ME source coding problem. We considered fixed-length codewords as a possible solution and analyzed its main parameters. The optimization procedure as well as system performance depends on number of symbols $M$ and cost ratio factor $\gamma=\frac{\beta_1}{\beta_0}$. Simulation results illustrate that significant energy saving, (which could be as high as \%70), is achieved. We also introduced variable-length coding scheme as another solution to the problem which was formulated as an integer program. It provides more flexibility in terms of codeword selection but it needs careful attention to selection of codewords for unique decodability for which we assumed prefix codes. Variable length coding generally out-performs fixed length coding and the difference was shown to be significant (up to 15\% for $M=8$).

Application of ME source coding in RFID was investigated by defining energy cost in RFID tags as the difference between harvested energy and tag consumed energy. The cost ratio factor $\gamma$ was also formulated in terms of RFID system parameters. A comparison of this method with current RFID standard, EPCglobal, proves a considerable energy saving.

\ifCLASSOPTIONcaptionsoff
  \newpage
\fi

\bibliographystyle{IEEEtran}
\bibliography{IEEEabrv,mylit}

\end{document}